\documentclass[journal,twocolumn]{IEEEtran}
\usepackage{amsfonts}
\usepackage[tbtags]{amsmath}
\usepackage{graphicx,subfigure}
\usepackage{multirow}
\usepackage{color}      
\usepackage{epsfig}
\usepackage{amssymb}
\usepackage{tipa}
\usepackage{bbm}

\long\def\comment#1{}

\newcommand{\beq}{\begin{equation}}
\newcommand{\eeq}{\end{equation}}
\newcommand{\beqno}{\begin{equation*}}
\newcommand{\eeqno}{\end{equation*}}

\newcommand{\bes}{\begin{split}}
\newcommand{\ees}{\end{split}}
\newcommand{\bdm}{\begin{displaymath}}
\newcommand{\edm}{\end{displaymath}}

\newtheorem{definition}{Definition}

\newcommand{\bd}{\begin{definition}}
\newcommand{\ed}{\end{definition}}

\newcommand{\bfx} {\mathbf{x}}  
\newcommand{\bfy} {\mathbf{y}}  

\newcommand{\bv}{\begin{vugraph}}
\newcommand{\ev}{\end{vugraph}}
\newcommand{\bi}{\begin{itemize}}
\newcommand{\ei}{\end{itemize}}
\newcommand{\ben}{\begin{enumerate}}
\newcommand{\een}{\end{enumerate}}

\newcommand{\bean}{\begin{eqnarray*} }
\newcommand{\eean}{\end{eqnarray*} }
\newcommand{\bea}{\begin{eqnarray} }
\newcommand{\eea}{\end{eqnarray} }

\newcommand{\ba}{\begin{array} }
\newcommand{\ea}{\end{array} }

\newcommand{\bfe}{\mathbf{e}}
\newcommand{\bfp}{\mathbf{p}}
\newcommand{\bfq}{\mathbf{q}}
\newcommand{\bfs}{\mathbf{s}}

\newcommand{\bfi}{\mathbf{i}}
\newcommand{\bfg}{\mathbf{g}}
\newcommand{\bfr}{\mathbf{r}}

\begin{document}
\title{Improved compression of network coding vectors using erasure decoding
and list decoding}

\author{Shizheng~Li ~\IEEEmembership{Student Member,~IEEE} and Aditya~Ramamoorthy,~\IEEEmembership{Member,~IEEE}
\thanks{The authors are with the Department of Electrical and Computer Engineering, Iowa State University, Ames, Iowa 50011, USA. Email: \{szli, adityar\}@iastate.edu}
\thanks{This research was supported in part by NSF grant
CNS-0721453. }}

\maketitle

\begin{abstract}
Practical random network coding based schemes for multicast
include a header in each packet that records the transformation
between the sources and the terminal. The header introduces an
overhead that can be significant in certain scenarios. In previous
work, parity check matrices of error control codes along with
error decoding were used to reduce this overhead. In this work we
propose novel packet formats that allow us to use erasure decoding
and list decoding. Both schemes have a smaller overhead compared
to the error decoding based scheme, when the number of sources
combined in a packet is not too small.
\end{abstract}
\section{Introduction}
In a multicast scenario, 
network coding can achieve maximum-flow-min-cut capacity.
It is shown in \cite{rm,hoMK06} that if each intermediate node
transmits random linear combinations of the incoming packets over
a large field,  the terminal can recover the source packets with high probability.
Under such a distributed randomized scheme, the terminals need to
know the transfer matrix. In \cite{chou03} it was shown that this
can be carried in the headers of the packets.
The header records the network coding vector, which consists of
the linear combination coefficients for the packet. The header length equals to the number of
source packets,
which is negligible when the packet length is large and the number
of sources is relatively small.

There are situations in which the packet overhead can be
significant. As noted in \cite{jafari09}, in sensor networks, the
number of sources is large and current sensor technology does not
allow transmission and reception of very large packets.
However, in many of these applications, the network topology is
such that the received packets at a terminal only consist of
combinations of a small or moderate number of sources. In
addition, the random network coding protocol can possibly be
appropriately modified to enforce the constraint that a received
packet contains combinations of only a few sources. This implies
that it may be possible to ``compress" the header size and reduce
the overhead. The idea of compressing coding vectors was first proposed in \cite{jafari09}, where a strategy using parity-check matrices of error
control codes was used. Under that scheme, the
overhead of each packet has length $2m$ if the maximum number of
packets being combined in the packet is $m$.

Suppose the total number of sources is $n$. As mentioned in \cite{jafari09}, the restriction on the number of
combined packets introduces $n-m$ zeros in each row of the
transfer matrix, which may affect the invertibility of the matrix.
The network topology in general will make the distribution of
zeros non-uniform and this makes the chance of losing rank becomes
larger. Therefore, the value of $m$ can not be too small.

\noindent{\underline{\it Main Contributions}} - In this work, we
propose improved schemes for the compression of network coding
vectors.

\noindent1) In the first scheme, we add an ID segment to the
header that records the IDs of the sources being combined in the
packet. This requires modifying the intermediate node operation
slightly but gives two main advantages: a) It allows us to convert
the problem at the terminal into one of decoding erasures (as
against decoding errors). The required header length becomes $m +
n/\log q$ (the base of the logarithm is two throughout the paper),
where $q$ is the field size. It is less than the overhead of the
error decoding based scheme ($2m$) when $m$ is not too small. b)
The protocol suggested in \cite{jafari09} to limit the number of
sources combined in a packet adds a counter to each packet for
tracking the number of sources that have been combined. However,
when combining two incoming packets, it is hard for the
intermediate node to know the number of source packets that will
be combined in the new packet because the sets of source packets
in the incoming packets may overlap. It can only obtain an
inaccurate upper bound by adding two counters together. Using our
proposed ID segment, the number of source packets being combined
in every coded packet can be accurately traced. 2) In the second
scheme, we propose a list-decoding based compression scheme (based
on error decoding like \cite{jafari09}) , whose overhead can be
made arbitrarily close to $m + O(\log n)/\log q$. In this scheme
the intermediate nodes remain oblivious to the fact the network
coding vectors are compressed. The lower overhead for this scheme
comes at the expense of higher decoding complexity (for the
header) at the terminal. \vspace{-1mm}
\section{Background and Related Work}
 Let $F_q$ denote a finite field with size $q$, where
$q$ is a power of two. Consider a network with $n$ sources, not
necessarily collocated. The $i^{th}$ source transmits a length-$N$
packet $\bfp_i\in F_q^N$. The packet contains two parts: $\bfp_i =
[\bfp_i^H|\bfp_i^M]$, where $\bfp_i^H \in F_q^h$ is the header and
$\bfp_i^M\in F_q^{N-h}$ is the actual message. The $i^{th}$ packet
received by a terminal is $\bfr_i=[\bfr_i^H|\bfr_i^M]$, where
$\bfr_i^H$ denotes the header and $\bfr_i^M$ denotes the coded
message. In \cite{chou03}, the header, $\bfp_i^H$ is designed to
be the $i^{th}$ row $\bfi_i$ of an $n$-by-$n$ identity matrix.
Thus, under random network coding, $\bfr_i^H$ contains the overall transformation from the sources to the
terminal for the coded message $\bfr_i^M$. The length of the
header $h=n$. Denote the vector of transformation coefficients by
$\bfq_i$.

In general, the entries of $\bfq_i$ could be all non-zero since
all sources could be combined. Under the assumption that at most
$m$ sources are combined, $\bfq_i$ contains at most $m$ non-zero
entries, which leads us to an error control coding based
compression \cite{jafari09}. Let $H$ be a parity check matrix of a
$(n,k,d)$ linear block code, where $d$ is the minimum distance
\cite{shulin}. In a channel coding setting, a codeword $\bfx$ such
that $\bfx H^T = 0$ is transmitted, and $\bfy = \bfx + \bfe$ is
received, where $\bfe$ denotes the error. The decoder computes the
syndrome (of length $n-k$) $\bfy H^T = \bfe H^T = \bfs$ and finds
the error pattern $\bfe$. As long as the actual Hamming weight of
$\bfe$, $wt(\bfe)\leq \lfloor(d-1)/2\rfloor$, $\bfe$ can be
recovered exactly. This can be done efficiently for codes such as
RS and BCH using the Berlekamp-Massey algorithm (BMA)
\cite{shulin}. Equivalently, we can reconstruct $\bfe$ (of
length-$n$) from $\bfs$ (of length $n-k$) and this can be viewed
as a method to compress a vector $\bfe$.
For an error pattern such that $wt(\bfe)\leq m$, to get a high
compression rate, we want $k$ to be as large as possible while the
minimum distance is $d$ and the code length is $n$. From the
Singleton bound \cite{shulin}, $k\leq n-d+1 = n-2m$ and the well
known RS codes achieve this with equality..

In the error-correction based compression scheme \cite{jafari09},
the header of the packet $\bfp_i$ injected in the network is
chosen to be $\bfp_i^H = \bfi_i H^T$. After random linear coding,
the $i^{th}$ received packet contains the header $\bfr_i^H =
\bfq_i H^T$. Note that the network coding vector $\bfq_i$ is a
length-$n$ vector with $wt(\bfq_i)\leq m$ and $\bfr_i^H$ is
available at the terminal. Thus, the problem of recovering
$\bfq_i$ is equivalent to error correction. Then the $n$ headers
can be stacked row by row, forming the $n$-by-$n$ transfer matrix.
The overhead is $h=n-k$ and the maximum number of sources
allowed to be combined in one packet is $m\leq \lfloor
h/2\rfloor$.
%

\vspace{-1.5mm}
\section{Erasure decoding based compression scheme}
 In channel coding, an erasure is defined to be an
error whose location is known by the decoder. For a linear block
code with minimum distance $d$, it can correct up to $d-1$
erasures. For BCH codes and RS codes, syndrome-based decoding and
the BMA still work after some minor modifications \cite{shulin}.
In the network coding vector compression scenario, if we know the
locations of non-zero elements in $\bfq_i$, we can allow $m$ to be
as large as $d-1\leq n-k$. Note that as long as we know which
source packets are combined in the packet of interest, we know the
locations of the non-zero elements.

\noindent\underline{{\it Proposed Solution.}} -  We add a bit
array of length-$n$ to the header $\bfp_i^H$ and call it {\it ID
segment}. At the $j^{th}$ source, only the $j^{th}$ position is
set to 1 and others are 0. At every intermediate node, when
several incoming packets are combined to form a packet for an
outgoing edge, the ID segment of the outgoing packet is the
bit-wise OR of the ID segments of the incoming packets. $\bfp_i^H$
also includes $\bfi_i H^T$ (of length $n-k$) as before. This
protocol is very easy to implement and every packet in the network
knows exactly which source packets are combined in it. The
$j^{th}$ element of $\bfq_i$ is non-zero if and only if the
$j^{th}$ bit in the ID segment of $\bfr_i^H$ is 1.
 As pointed out in the introduction, if we want to
limit the number of source packets being combined by network
protocol, this information is important for the intermediate
nodes. The terminal receives the ``syndrome" $\bfq_i H^T$ and
knows the locations of the ``errors". By erasure decoding, it can
recover $\bfq_i$ as long as $wt(\bfq_i)\leq m=n-k$.

The length of the ID segment in terms of
symbols is $n/\log q$. The total overhead is $n-k+n/\log q$. If
$m$ is fixed, the overhead for the scheme in
\cite{jafari09} is $2m$ and the overhead for our erasure decoding
scheme is $m+n/\log q$. Thus, if $m$ is not too small, our
proposed scheme has less overhead.


{\it Example 1}. Suppose $n=50, q = 2^8, m = 15$. Under error
decoding scheme, a $(50, 20)$ RS code is required and the overhead
is $30$ bytes. Under erasure decoding scheme, a $(50, 35)$ RS code
is required and the overhead is $22$ bytes, a saving of 26\%.
According to the current ZigBee standard \cite{Zigbee}, the packet
size is $128$ bytes.

{\it Example 2}. Suppose $n=255, q = 2^8, m = 150$. No code has
minimum distance 301 with code length 255. Under error decoding
the network coding vector cannot be compressed and the overhead
$h=n=255$. Under erasure decoding scheme, a $(255, 105)$ RS code
can be used and $h=182$.


A reviewer has pointed out that if one uses a bit-array to record
the IDs of the sources, then there is an alternative scheme that
does not require decoding at the terminals. Basically, every node
keeps track of the coefficients and the ID's and combines them so
that the net transformation is available at the terminals without
decoding. However, such a scheme requires the intermediate nodes
to scan the headers of the incoming packets to locate the
corresponding coefficients that need to be combined (in addition
to performing a bitwise OR in the ID array). This solution
increases the processing complexity at the intermediate nodes.
Our proposed approach can be viewed as an alternate solution to
this problem. The correct choice would depend upon the
capabilities of the sensor nodes and the application requirements.


\vspace{-1mm}
\section{List decoding based compression scheme}
In this section, we show that the overhead of the strategy based
on error decoding (such as \cite{jafari09}) can be reduced by using list
decoding at the terminal. It does not require the decoder to know
the error locations so we need not add the ID segment in the
header. Furthermore, the intermediate nodes simply perform linear
combination on the header, i.e., it is oblivious to the fact the
network coding vectors are compressed. In the channel coding
scenario, given the received word $\bfy = \bfx + \bfe$, the
decoder tries to find a codeword $\bfx$ within Hamming distance
$t_0 \triangleq \lfloor(d-1)/2\rfloor$ of $\bfy$. As long as
$wt(\bfe)\leq t_0$, the decoder will find a unique $\bfx$ and the
decoding is successful. When $wt(\bfe)>t_0$, there is no guarantee
that the decoder will succeed. This is the scenario in which the
notion of list decoding is useful. The list decoding problem can
be stated as follows.

\textit{Problem 1. Given a received word $\bfy = \bfx + \bfe $,
find the list of all codewords $\bfx$'s within Hamming distance
$t>t_0$ of $\bfy$.}

As long as $wt(\bfe)\leq t$, the actual codeword $\bfx$ will
appear in the list. The list decoding problem has been solved to
some extent  (see \cite{Gmono07} for a survey). Efficient list
decoding algorithms with polynomial sized lists for RS codes up to
a radius of $(t=n - \sqrt{nk})$ are known. The class of folded RS
codes \cite{Gmono07} can be decoded arbitrarily close to the
Singleton bound, i.e., $t$ can be close to $n-k$, though this is
possible only with very large alphabets. In order to apply list
decoding to our problem, we propose a packet header for the
$i^{th}$ source packet that consists of $\bfi_i H^T$ and some side
information. Note that at the terminal, we obtain the syndrome
$\bfs=\bfe H^T = \bfq_i H^T$
of network coding vector
.Therefore, the problem can be stated as follows.
\par\textit{Problem 2. Find the list of all possible error pattern $\bfe$'s such that $\bfe H = \bfs$ and $wt(\bfe)\leq t$, where $t>t_0$.}

We present a problem transformation such that all list decoding
algorithms for problem 1 can be used to solve problem 2. Given
$\bfs=\bfe H^T$, we can find an arbitrary $\bfy$ such that
$\bfs=\bfy H^T$, then use this $\bfy$ as input to problem 1 and
get the list of $\bfx$'s as an output, then $\bfe = \bfx + \bfy$
form the list of $\bfe$'s. Such $\bfy$ can be chosen easily.
Recall that the parity check matrix $H$ of a $(n,k)$ code has rank
$(n-k)$ and there exist $(n-k)$ columns in $H$ that are linearly
independent. Let the elements of $\bfy$ that correspond to these
columns  be unknowns and other $k$ elements be zero. Note if a RS
code is used, we can choose any $k$ elements in $\bfy$ to be zero.
The system of equations $\bfs = \bfy H^T$ has $(n-k)$ unknowns and
$(n-k)$ linearly independent equations, from which $\bfy$ can be
determined. Next, we prove that the above transformation solves
problem 2 correctly. Suppose the resultant list of problem 2 is a
set $L_1$ and the list obtained by using our transformation is a
set $L_2$. We need to show $L_1 = L_2$. First, if $\bfe \in L_2$,
since $\bfe = \bfx + \bfy$ and $\bfx$ and $\bfy$ differ at most
$t$ positions, $wt(\bfe)\leq t$ and $\bfe H^T = \bfx H^T + \bfy
H^T = 0 + \bfy H^T = \bfs$, then $\bfe\in L_1$. Second, if
$\bfe\in L_1$, there exists an $\bfx = \bfy+\bfe$ such that $\bfx
H^T = \bfy H^T + \bfe H^T = 0$ and since $wt(\bfe)\leq t$,
$\Delta(\bfx,\bfy) \leq t$ ($\Delta(\cdot)$ denotes Hamming
distance), this means $\bfx$ is a codeword within Hamming distance
$t$ of $\bfy$, then $\bfx$ is on the list of the output of problem
1. Thus $\bfe\in L_2$.


Note that so far we have only found a list of possible error
patterns. In practice we need to find the unique error pattern as
the decoded network coding vector. The small amount of side
information included in the header is useful here. The side
information generation problem was solved in \cite[Theorem
2]{listsideinfo03}. It is a hash function based algorithm to
select a message in a candidate set and works no matter we are
facing problem 1 or problem 2. Note that in our compression
problem, the message space is all possible network coding vectors
and the size is $q^n$. The side information at
the terminal should contain \cite[Lemma 1]{listsideinfo03} (i)
$\bfq_i\cdot \bfg_r$, where $\bfq_i$ is the actual ``message"
(network coding vector), $\bfg_r$ is a randomly chosen column of
the generator matrix of a low rate RS code (which is different
from the one used to generate the syndrome)
and $\cdot$ denotes inner product, and (ii) the random number $r$.
Denote the list of candidates to be
$\{\bfq_i^1,\ldots,\bfq_i^L\}$. The terminal knows the RS code a
priori and computes $\bfq_i^j \cdot \bfg_r$ for every $j$ and
finds $j^*$ such that $\bfq_i^{j*}\cdot \bfg_r = \bfq_i \cdot
\bfg_r$ . Since the actual $\bfq_i$ is in the list, such a $j^*$
exists. It was shown in \cite[Theorem 2]{listsideinfo03} that as
long as $O(\log n) + O(\log L) + O(\log(1/P_f))$ bits of side
information are provided, the probability that $j^*$ is not unique
is less than $P_f$. The basic idea behind this is that for two
codewords of a RS code with very large minimum distance, the
probability that the symbols at a random chosen position $r$ are
equal is very small. The list size $L$ is polynomial with $n$.
Thus, the amount of side information needed is $O(\log n)$ and
$P_f$ is the probability of failure to find a unique output. In
order to obtain the side information at the terminal, we include
$\bfi_i\cdot \bfg_r$ in the header of the $i^{th}$ source packets
and the intermediate nodes perform linear combination on it, so
that the terminal receives $\bfq_i\cdot \bfg_r$. We can let the
session ID to be the random number $r$ and available to the
sources and terminals so that $r$ does not need to be transmitted
over the network.

The list decoding based scheme incurs an overhead of length
$m+O(\log n)/\log q$ and allow the number of source packets being
combined to be $m$. It has smaller overhead size than erasure
decoding based scheme. However, as mentioned before, in order to
approach the list decoding capacity, the field size needs to be
large and the decoding algorithm becomes more complicated. If we
use ordinary RS codes and the efficient decoding algorithms that
corrects up to $n-\sqrt{nk}$ errors to compress network coding
vector, the overhead length will be $2m-m^2/n + O(\log n)/\log q$.
Usually this will be less than the overhead of error decoding
based scheme but greater than erasure decoding based scheme.

{\it Example 3}. Suppose $n=255, q=2^8, m=86$. We use a
$(255,112)$ RS code.
The syndrome length is 143 and the side
information length is $\lceil 30/8\rceil$ for $P_f=0.0001$,
so $h=147$. $h$ equals $172$ or 118 for error or erasure decoding
respectively. \vspace{-1mm}
\section{conclusion}
We proposed erasure decoding based and list decoding based
approaches to improve the compression of network coding vectors.
Table \ref{tab:comp} compares the overheads of the various
schemes. For moderate or large value of $m$, that may be necessary
to support the multicast rate, both schemes have less overhead
than the error decoding based scheme. Our investigation reveals
that the list decoding based scheme has a lower overhead with
respect to the erasure coding based scheme, when capacity
achieving codes are used. However, from a practical perspective,
the erasure coding scheme offers the best tradeoff between
overhead and implementation complexity.
\begin{table}[h]
\begin{center}
\caption{\label{tab:comp} Comparison of three schemes for the same
$m$.}
\begin{tabular}{|c|c|c|}
\hline & Header format & Header length\\ \hline Error& Syndrome &
$2m$\\ \hline
 \multirow{2}{*} {Erasure} & Syndrome  & {$m+n/\log q$}\\
& + ID segment & \\
 \hline \multirow{2}{*}   {List} &
Syndrome & $m+O(\log n)/\log q$ \\ & + side information & or
$2m-m^2/n + O(\log n)/\log q$\\ \hline
\end{tabular}
\end{center}
\end{table}
\vspace{-3mm}


\bibliographystyle{IEEEtran}
\bibliography{CompNCV}

\begin{thebibliography}{1}
\providecommand{\url}[1]{#1}
\csname url@samestyle\endcsname
\providecommand{\newblock}{\relax}
\providecommand{\bibinfo}[2]{#2}
\providecommand{\BIBentrySTDinterwordspacing}{\spaceskip=0pt\relax}
\providecommand{\BIBentryALTinterwordstretchfactor}{4}
\providecommand{\BIBentryALTinterwordspacing}{\spaceskip=\fontdimen2\font plus
\BIBentryALTinterwordstretchfactor\fontdimen3\font minus
  \fontdimen4\font\relax}
\providecommand{\BIBforeignlanguage}[2]{{%
\expandafter\ifx\csname l@#1\endcsname\relax
\typeout{** WARNING: IEEEtran.bst: No hyphenation pattern has been}%
\typeout{** loaded for the language `#1'. Using the pattern for}%
\typeout{** the default language instead.}%
\else
\language=\csname l@#1\endcsname
\fi
#2}}
\providecommand{\BIBdecl}{\relax}
\BIBdecl

\bibitem{rm}
R.~Koetter and M.~M\'{e}dard, ``{An Algebraic Approach to Network Coding},''
  \emph{IEEE/ACM Trans. on Netw.}, vol. 11, no. 5, pp. 782--795, 2003.

\bibitem{hoMK06}
T.~Ho, M.~Medard, R.~Koetter, D.~Karger, M.~Effros, J.~Shi, and B.~Leong, ``{A
  Random Linear Network Coding Approach to Multicast},'' \emph{IEEE Trans. on
  Info. Th.}, vol. 52, no. 10, pp. 4413--4430, 2006.

\bibitem{chou03}
P.~A. Chou, Y.~Wu, and K.~Jain, ``{Practical Network Coding},'' in \emph{41st
  Allerton Conference on Communication, Control, and Computing}, 2003.

\bibitem{jafari09}
M.~Jafari, L.~Keller, C.~Fragouli, and K.~Argyraki, ``Compressed network coding
  vectors,'' in \emph{Proc. IEEE Int. Symp. Inf. Theory}, Jun. 2009.

\bibitem{shulin}
S.~Lin and D.~J. Costello, \emph{Error control coding: fundamentals and
  applications}.\hskip 1em plus 0.5em minus 0.4em\relax Prentice Hall, 2004.

\bibitem{Zigbee}
E.~Callaway, P.~Gorday, L.~Hester, J.~Gutierrez, M.~Naeve, B.~Heile, and
  V.~Bahl, ``Home networking with ieee 802.15.4: a developing standard for
  low-rate wireless personal area networks,'' \emph{Communications Magazine,
  IEEE}, vol.~40, no.~8, pp. 70 -- 77, aug 2002.

\bibitem{Gmono07}
V.~Guruswami, \emph{Algorithmic Results in List Decoding}.\hskip 1em plus 0.5em
  minus 0.4em\relax Now Publishers, 2007.

\bibitem{listsideinfo03}
V.~Guruswami., ``List decoding with side information,'' in \emph{Proceedings of
  Computational Complexity}, 2003.

\end{thebibliography}

\end{document}